 \documentclass[
prl,twocolumn,showpacs, amsmath, amssymb, floatfix,
nofootinbib,wrapfloat byrevtex]{revtex4}

 \usepackage{amsmath,mathrsfs,upgreek,graphicx,epstopdf,placeins,float}
 \usepackage{booktabs}
 \usepackage{multirow}

 \newcommand{\beq}[1]{\begin{equation}\label{#1}}
 \newcommand{\eeq}{\end{equation}}
 \newcommand{\bea}[1]{\begin{eqnarray}\label{#1}}
 \newcommand{\eea}{\end{eqnarray}}
 \newcommand{\m}{{\rule[1.5pt]{2pt}{0.5pt}}}
 \newcommand{\mm}{{\rule{2pt}{0.5pt}}}
 \newcommand{\cddot}{{\cdot\cdot}}

\begin{document}

 \title{Schwarzschild Fuzzball and Explicitly Unitary Hawking Radiations}
 \author{Ding-fang Zeng}
 \email{dfzeng@bjut.edu.cn}
 \affiliation{Institute of Theoretical Physics, Beijing University of Technology, China, Bejing 100124}
 \begin{abstract}
We provide a fuzzball picture for Schwarzshild black holes, in which matters and energy consisting the hole are not positioned on the central point exclusively but oscillate around there in a serial of eigen-modes, each of which features a special level of binding degrees and are quantum mechanically possible to be measured outside the horizon. By listing these modes explicitly for holes as large as $6M_\mathrm{pl}$, we find that their number increases exponentially with the area. Basing on this picture, we present a simple but explicitly unitary derivation of hawking radiations. 
 \end{abstract}
 \maketitle

The horizon and central singularity are two key ingredients of general relativistic black holes, either from  observational \cite{cardoso2017} or from pure theoretical \cite{mathur2009} aspects. They are also birth-lands of many radical proposition and exciting progresses in quantum gravitation researches, typically the information missing puzzle \cite{hawking1976, hawking1975cmp, Mathur0909, polchinski2016} and the Anti-de Sitter/Conformal Field Theory correspondence \cite{maldacena1997} or more generally the gauge/gravity duality (AdS/CFT here after). Although initiative researchers such as L. Susskind, basing on general ideas of gauge/gravity duality and special picture of string theories \cite{DasMathur9601, MaldacenaSusskind9604, StromingVafa9601, Mathur9706, MT0103, LuninMathur0109, LuninMathur0202, LuninMaldacenaMaoz0212, GMathur0412, mathur0502, Jejjala0504, KST0704, Mathur0706}, claims that the war between him and S. Hawking has finished already \cite{blackholewar}, new ideas on the information puzzle's reformulation and resolution continue to appear endlessly, ranging from the famous AMPS observation of firewall paradoxes \cite{fireworksAMPS2012,fireworksAMPS2013} and the ER=EPR proposition \cite{EREPR}, to various nonlocal/entanglement \cite{stojkovic1401, stojkovic1503, stojkovic1601, bardeen1706, halyo1706, stojkovic0609, sunYi1102, KMY1302, zhangBc1305, caiQingyu1608, hopeiming1609} revision believed being ignored in hawking's original calculation, and to totally new mechanisms for black holes to save information \cite{hawking2016}, although in challenges \cite{bousso1706,giddings1706}.

{\em Basic ideas} The general idea of gauge/gravity duality that microstates of black holes could be explained in terms of  lower dimensional gauge field theories brings us misunderstandings that, the information of the black holes is stored locally in their near horizon region. However, even in the most well understood fuzzball picture of string theories \cite{Mathur9706, MT0103, LuninMathur0109, LuninMathur0202, LuninMaldacenaMaoz0212, GMathur0412, mathur0502, Jejjala0504, KST0704, Mathur0706}, S. D. Mathur, et al tell us that for large classes of asymptotically AdS black holes constructible from or related with special D1-D5 brane configurations, the information carriers are distributed across the whole region covered by the horizon surface. For more general black holes, especially the Schwarzschild ones, the string theory still finds no way to give the relevant information saving mechanism a concrete explanation. We considered in ref.\cite{dfzeng2017} a possibility that, matters inside the Schwarzschild black holes, which we call Schwarzschild contents in this paper, are not positioned on the central point of the hole statically but are experiencing periodical motion of collapsing, collapsing overdone to the other side and collapsing again during which the radial mass profile preserves continuously. We argue that it is just this radial mass profiles' diversity, chosen at arbitrary given times $\tau=\tau_0$, with the future determined by Einstein equation, that leads to the microstates' multiplicity of black holes. The inner metric of these holes when written in the co-moving observer's proper time has the form
\bea{}
&&\hspace{-5mm}ds^2=-(h^{-1}\frac{\dot{m}^2}{m'^2}+1)d\tau^2+h^{-1}dr^2+r^2d\Omega_2^2
\\
&&\hspace{-5mm}h=1-\frac{2Gm(\tau,r)}{r},~r<r_0\equiv2Gm_\mathrm{total}
\eea
By looking $m(\tau_0,r)$ as independent coordinate and introducing a wave functional $\Psi[m(\tau_0,r)]$ to denote the amplitude the hole being at profile $m(\tau_0,r)$, we establish in ref.\cite{dfzeng2017} a functional differential equation controlling the form of $\Psi[m(\tau_0,r)]$ through quantisations of the Hamiltonian constraint of the system, thus translate the question of black hole microstates' definition and counting a functional eigenvalue problem. However, due to complexes of the functional differential equation, we get only rough estimations for the eigen-state of $1$- and $2$-$M_\mathrm{pl}$ mass black holes. The purpose of this work is to provide an alternative definition for this functional problem and an almost exact thus more convincing proof of the microstate number's exponential area law. Basing on this proof, we will also give a hamiltonian thus explicitly unitary derivation for hawking radiations.

\begin{figure}[ht]
 \includegraphics[scale=1,clip=true,bb=2 35 237 110]{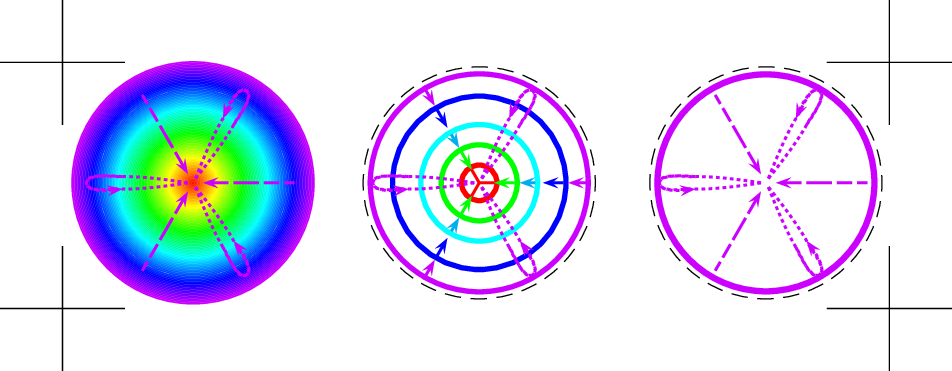}
 \caption{Left, practical Schwarzschild contents may have continuous, dynamically evolving radial mass distribution profile. Middle, each radial mass profile can be considered a direct product of many concentric dynamically evolving mass shells. We will show that the number of this layering scheme is finite and countable. Right, the most simple layering scheme is that the whole content consists of just one oscillating shell. We will show by simple quantum mechanics that this shell could only be at some discrete quantum states, each of which corresponds to a classic shell released from some special initial radius inside the horizon surface.}
 \label{figLayerIdea}
 \end{figure}

The classic picture behind our micro-state definition and counting is shown in FIG.\ref{figLayerIdea} schematically. Just as was done in \cite{dfzeng2017}, we will still focus on black holes consisting of zero pressure dusts for simplicity. The advantage of this doing is that, we can easily prove that the co-moving observer's geodesic motion $\{u^0=1,g_{\mu\nu}u^\mu u^\nu=-1\}$ follows directly as part of Einstein equation $G_\mu^{~\nu}=8\pi GT_\mu^{~\nu}$, i.e. $\frac{G_0^{~1}}{G_0^{~0}}\stackrel{e.e.}{=}\frac{u^1}{u^0}\stackrel{g.m.}{=}\frac{\dot{m}}{m'}$. But different from \cite{dfzeng2017}, in this time we will decompose the radial mass distribution profile into several concentric shells at the first beginning. We will show that the number of this shell partition, as well as their quantum state are both countable, with the latter equals to $e^{kA/4G}$. In the mostly simple layering scheme, all contents of the hole are concentrated in one shell of dusts, the equations of motion, looking from exterior observers which use time $t$ and feel a Schwarzschild geometry  v.s. interior observers which use $^{\scriptscriptstyle\prime}\!t$ and feel a Minkowskian geometry, can be easily written as
\beq{}
\Big\{\begin{array}{l}h\dot{t}=\gamma\Leftarrow\ddot{x}^0+\Gamma^0_{\mu\nu}\dot{x}^\mu\dot{x}^\nu\\h\dot{t}^2-h^{-1}\dot{r}^2=1\end{array}
\!\!,~\Big\{\begin{array}{l}^{_\prime}\!\dot{r}=\dot{r},~h=1-\frac{2Gm}{r}\\
^{\scriptscriptstyle\prime}\!\dot{t}^2-^{\scriptscriptstyle\prime}\!\!\dot{r}^2=1\end{array}
\label{classicEom}
\eeq
\beq{}
\mathrm{or}~\Big\{\begin{array}{l}\dot{r}^2=\gamma^2-h\\
\dot{t}^2=\gamma^2h^{-2}
\end{array},~
\Big\{\begin{array}{l}^{\scriptscriptstyle\prime}\!\dot{r}^2=\gamma^2-h,~h=1-\frac{2Gm}{r}\\
^{\scriptscriptstyle\prime}\!\dot{t}^2=\gamma^2-h+1
\end{array}
\label{classicEomSimp}
\eeq
where $\gamma$ is an integration constant equaling to the value of $h$ on $r=r_\mathrm{rel}$ where the shell is released from static and allowed to freely moving under self-gravitations. If the shell is released from outside the horizon, $\gamma$ will be real and less than $1$. But we will focus on cases where the shell is released from inside the horizon, so that $\gamma$ is pure imaginary and $\gamma^2<0$. Talking about this motion inside horizons is meaningful because we could have observers on the central point around which the space time is simply minkowskian before the shell arrives onto. While as the shell arrives onto, its radial speed equals to that of light $\frac{d^{\scriptscriptstyle\prime\!\!}r^2}{d^{\scriptscriptstyle\prime}\!t^2}\xrightarrow{^{\scriptscriptstyle\prime\!\!}r\rightarrow0}1$, so it cannot be stopped there but have to go across that point and making oscillations there. 

{\em Microstate definition and counting} For the singular point across oscillation, quantum descriptions are necessary but straightforward. For the most simple one-shell contents layering scheme, we need only introduce a wave function $\Psi[r]$ to denote the probability amplitude the mass shell could be measured at $r$, and quantise the equation of motion $\dot{r}^2=\gamma^2-h$ or $^{\scriptscriptstyle\prime}\!\dot{r}^2=\gamma^2-h$ directly. The result is
\beq{}
[-\hbar^2\partial_r^2-m^2(\gamma^2-h)]\Psi=0,~
\gamma^2\equiv h[r_\mathrm{rel}]\in(-\infty,0)
\label{schrodingerEqOneLayer}
\eeq
\beq{}
\mathrm{or}~
[\partial_{\hat{r}}^2-\beta^2+\frac{\hat{a}}{\hat{r}}]\Psi(\hat{r})=0,1<-\gamma^2\!+\!1\equiv\beta^2
\label{schrodingerEqOneLayerSimp}
\eeq
\beq{}
\hat{r}\equiv rm\hbar^{-1},~\hat{a}\equiv 2Gmm\hbar^{-1}
\eeq
It can be easily verified that, the vanishing at origin, square integrable solution to this equation of motion  exists only when $\frac{\hat{a}}{2\beta}=1,2,\cdots,n$. This is almost the same as the simple hydrogen atoms. However, due to the condition that the shell is released from inside the horizon thus $\gamma^2=h(r_\mathrm{rel})<0$, we must have $1<\beta$. This constrains the allowed wave function to be, $L^1_{q-1}(x)$ here is the first order associated Legendre polynomial, 
\beq{}
\Psi=\Psi_\beta(\hat{r})=e^{-\beta\hat{r}}\beta\hat{r} L_{q-1}^1(2\beta\hat{r})
\label{psiAllowedOneLayer}
\eeq
\beq{}
1\leqslant\beta,~\frac{2Gmm\hbar^{-1}}{2\beta}\equiv q=1,2,\cdots,q_\mathrm{max}
\label{QConeLayer}
\eeq
That is, for a sphere shell of given mass, the number of allowed quantum wave functions corresponding to classic oscillation modes released statically from inside the horizon is finite, equals to the maximal integer no larger than $Gmm\hbar^{-1}$ or symbolically
\beq{}
q_\mathrm{max}=\mathrm{Floor}[Gmm\hbar^{-1}]
\label{nAllowedOneLayer}
\eeq

\begin{figure}[ht]
\includegraphics[scale=0.5]{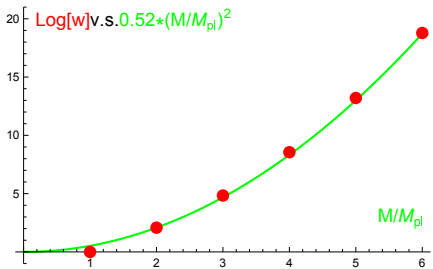}
\caption{The logrithmic value of the number of microstate of Schwarzschild black holes with masses less than or equal to 6-$M_\mathrm{pl}$.}
\label{figNumberBHs}
\end{figure}

While for the more interesting case where the whole Schwarzschild contents are consisting of several layer of different mass shells $m=\{m_1,m_2,\cdots,m_\ell\}$. In the case when all these shells do not cross each other, the wave function of the whole system can be written as $\Psi^1\otimes \Psi^2\otimes\cdots\otimes\Psi^{\ell}$. While if shell crossing occurs \cite{Yodzis1973, Yodzis1974}, we will need some symmetrisation procedure on this direct product to get wave functions of the system which have similar physical interpretations. But as long as the microstate number counting is concerned, such a consideration is not needed. Denoting the total mass inside the $i$-th shell, including the $i$-th shell itself, with $M_i\equiv\sum_{j=1}^i m_j$, and repeating the calculations across \eqref{classicEom}-\eqref{nAllowedOneLayer}, we will find that for each mass shell $m_i$, the corresponding wave function are simply
\beq{}
\Psi^i_m\!=\!\Psi^i_{m\beta_i}\!(\hat{r})\!=\!e^{-\beta_ir}\beta_i\hat{r}L^1_{q_i-1}(2\beta_i\hat{r}),\hat{r}=rm_i\hbar^{-1}
\label{psiAllowedMultiLayer}
\eeq
\beq{}
\frac{2GM_im_i}{\beta_i\hbar}\!\equiv\!q^m_i\!=\!1,2,\cdots,q^m_{i\mathrm{max}}\!\equiv\!\mathrm{Floor}[\frac{GM_im_i}{\hbar}]
\label{QCmultiLayer}
\eeq
So the total number of microstate allowed by this layering scheme $m$, and by the whole Schwarzschild contents equals to respectively
\beq{}
w_m=\prod_{i=1}^\ell q^m_{i\mathrm{max}},~w=\sum_m w_m
\eeq
This way, the question of black hole microstates' counting becomes a simple mass/energy layering partition and the corresponding quantum number listing. For a 2-$M_\mathrm{pl}$ mass black holes, we have only $3$ methods to partition the mass/energy contents into layers that lead to distinguished quantum state. The results are presented in TABLE.\ref{shelling2mpl}. While TABLE.\ref{shelling3mpl} lists all the quantum numbers allowed by a 3-$M_\mathrm{pl}$ mass black holes. For more large black holes up to 6-$M_\mathrm{pl}$ masses, listing out and counting up their quantum numbers one by one are also possible using computer programs. However,  as the black hole becomes even larger, the number of quantum states allowed by their contents increases exponentially with their horizon area. We plot the results in FIG.\ref{figNumberBHs} explicitly, from which we easily see that the entropy of the system
\beq{}
S\!\equiv\!k_{\!_B}\mathrm{Log}[w]\!=\!\frac{k'_{\!_B} A}{4G},~k'_{\!_B}\!=\!\frac{0.52k_{\!_B}}{4\pi},A\!=\!16\pi G^2M^2
\eeq
Except for a numeric factor of $\frac{0.52}{4\pi}\approx\frac{1}{8\pi}$, this yields perfectly the area law of  Bekenstein-Hawking formulas. We will not distinguish $k_{\!_B}$ and $k'_{\!_B}$ in the following and will denote them ambiguously as $k$. We note here that this result is only for 4-dimensional black holes. Continuations to other-dimensions are possible but nontrivial. For example, in 5-dimensions, the function $h$ appearing in  \eqref{schrodingerEqOneLayer} should be changed to $1-\frac{2G_{\!5}m}{r^2}$. This change will bring us to the very trouble Calogero problem \cite{GTV2012}, which is still not understood clearly in quantum mechanics but necessary for deriving  conditions like \eqref{QConeLayer} and \eqref{QCmultiLayer}. 

\begin{table}
\begin{tabular}{cccc}
\hline
\{$m_i$\}/$M_\mathrm{pl}$&\{2\}&\{1,1\}&\{$\frac{3}{2}$,$\frac{1}{2}$\}
\\
\{$M_i\!\equiv\!\sum m_{j\leqslant i}$\}&\{2\}&\{1,2\}&\{$\frac{3}{2}$,2\}
\\
$GM_im_i$&\{4\}&\{1,2\}&\{$\frac{9}{4}$,1\}
\\
$\frac{GM_im_i}{(1\leqslant)\beta}\in\mathbb{Z}$&\{1,2,3,4\}&\{(1),(1,2)\}&\{(1,2),(1)\}
\\
num.States&4&$1\cdot2$&$2\cdot1$
\\
\hline
\end{tabular}
\caption{The layering scheme and corresponding quantum states of a $2M_\mathrm{pl}$ mass black holes. Such a black hole can be layered into (i) one shell of total mass $2M_\mathrm{pl}$ or (ii) two shells of equal mass $1M_\mathrm{pl}+1M_\mathrm{pl}$ or (iii) two shells of unequal mass $\frac{3}{2}M_\mathrm{pl}+\frac{1}{2}M_\mathrm{pl}$. Other layering scheme such as $\{M_\mathrm{pl}+\epsilon,M_\mathrm{pl}-\epsilon\}$ with $0<\epsilon<\frac{M_\mathrm{pl}}{2}$ would not lead to radial quantum numbers different from listed above. While layerings such as $\{\frac{3M_\mathrm{pl}}{2}+\epsilon,\frac{M_\mathrm{pl}}{2}-\epsilon\}$ with $0<\epsilon<\frac{M_\mathrm{pl}}{2}$ break conditions that $\frac{GM_2m_2}{(1\leqslant)\beta}\in\mathbb{Z}$. So the total number of all possible quantum state is $8$.}
\label{shelling2mpl}
\end{table}

\begin{figure}[ht]
\includegraphics[scale=0.43]{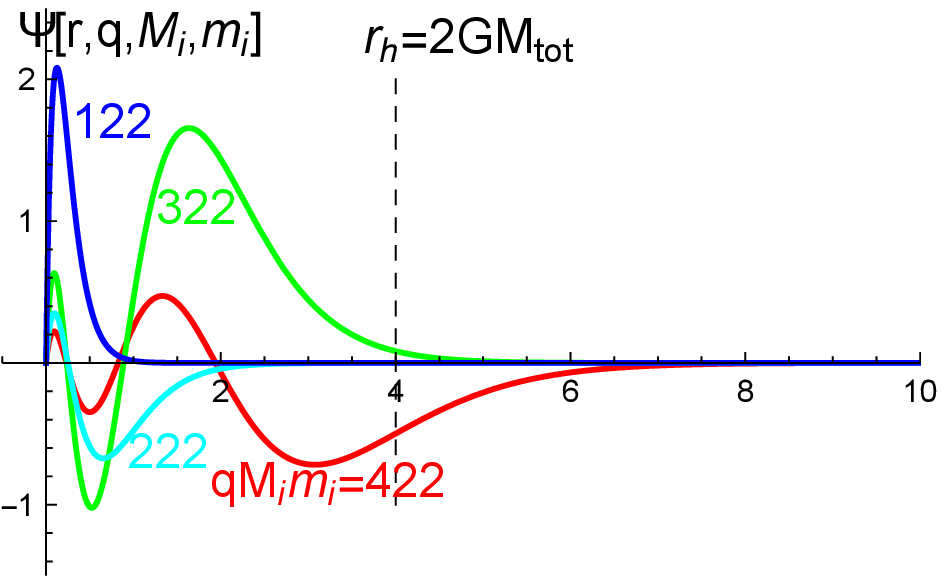}
\includegraphics[scale=0.43]{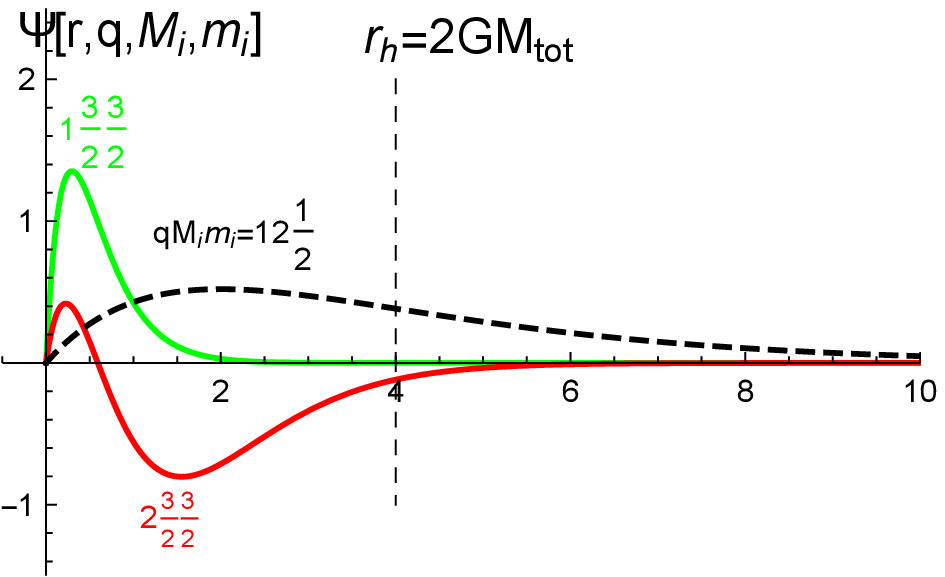}
\caption{Typical wave functions corresponding to eigen-modes of motion executed by the mass/energy contents of a $2M_\mathrm{pl}$ mass black holes. In the left panel, contents of the hole are concentrated in one mass shell which has only four possible quantum states. While in the right panel, the contents are distributed in two layer of shells each with mass $\frac{3M_\mathrm{pl}}{2}$ and $\frac{M_\mathrm{pl}}{2}$ respectively. In this latter case, the wave function of the system is the direct product of outside layer's $\Psi[r,12\frac{1}{2}]$ and inside layer's $\Psi[r,2\frac{3}{2}\frac{3}{2}]$ or $\Psi[r,1\frac{3}{2}\frac{3}{2}]$. }
\label{figEigenState}
\end{figure}

So, let us come back to the physic meaning of the black hole contents' wave function $\Psi_{\beta\propto1/q}[r]$ themselves. We plot in FIG.\ref{figEigenState} six typical wave functions of this type for a 2-$M_\mathrm{pl}$ mass Scharzschild black hole. From the figure, we firstly see that either in the one  or two layer partition case, we always have nonzero probabilities to find the contents  being outside the horizon. The horizon, behaves only approximately as the boundary of contents distribution. That is, it only requires that maximal values of the wave function occur inside it. From this aspects, our pictures are almost a quantum mechanic version of the string theory fuzzy balls \cite{Mathur9706, MT0103, LuninMathur0109, LuninMathur0202, LuninMaldacenaMaoz0212, GMathur0412, Jejjala0504, KST0704, Mathur0706, Mathur0909}. However, we get this picture basing on standard general relativity and simple quantum mechanics, instead of metaphysic ideas such as extra-dimension or supersymmetry et al.
The second point we need to emphasise in FIG.\ref{figEigenState} is about the physic meaning of quantum numbers $q^m_i$ and $\beta^m_i\equiv GM_im_i/q^m_i\hbar$, the superscript $m$ here denotes the mass shell partition scheme $m=\{m_1,m_2,\cdots\}$. From the figure, we easily see that for more larger $\beta^m_i$ or more smaller $q^m_i$, the global maximal point of the wave function occurs more close to the central point. The relevant microstate corresponds to classical shells binding more compactly to the central point. When we make direct product with these $\beta^m_i$ or $q^m_i$s, what we get is nothing but a quantitative characterising of the system's binding degree status. Due to differences between the mass/energy deficit during the bound state formation, an $r_h=2G\cdot2M_\mathrm{pl}$ black hole in e.g., state $\Psi[r,1,\frac{3}{2},\frac{3}{2}]\otimes\Psi[r,1,2,\frac{1}{2}]$ may possibly follow from collapsing of more masses than an $r_h=2G\cdot3M_\mathrm{pl}$ one in, e.g, state $\Psi[r,9,3,3]$.

\begin{figure}[ht]
\includegraphics[scale=1,clip=true,bb=2 30 220 310]{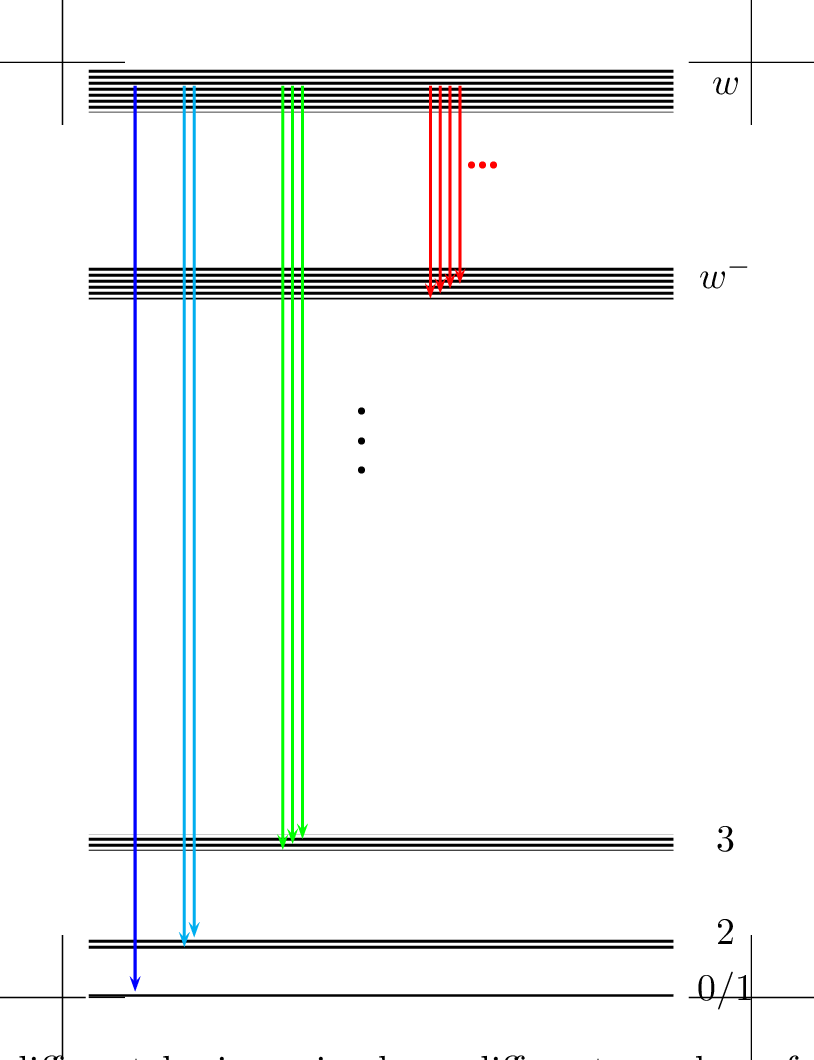}
\caption{Black holes of different horizon size have different number of microstates. In conventional hawking radiations, the initial state is unknown to be in which of the $w$ possibilities. While the final state has totally $w^-+w^{--}+\cdots+1+1$ possibilities. $w^-\equiv w-1$, $w^{--}\equiv w-2$, $+1+1$ because both the zero-mass vacuum and the 1-$M_\mathrm{pl}$ mass black hole are one-time degenerated.}
\label{figDecayLevels}
\end{figure}

{\em Explicitly unitary hawking radiations} By above picture of micro-state explanation, hawking radiations of a black hole are nothing but spontaneous decays of its contents' binding state. However, in pure geometric derivations of hawking radiations, effects of the initial state's difference are ignored so must be averaged in microscopic derivations. Referring to FIG.\ref{figDecayLevels}, we easily see that the probability for a mass $b$ black hole to spontaneously emit a mass $\Delta$ hawking particle and becomes a mass $b-\Delta$ hole is proportional to [the parameter $(8\pi kGb)^{-1}$ will be denoted by an effective temperature $kT$]
\beq{}
P_{\!_\Delta}\!=\!\frac{e^{k\pi G4(b-\Delta)^2}}{\frac{(e^{k\pi G4b^2}-1+1)e^{k\pi G4b^2}}{2}}\!\stackrel{\Delta\ll b}{=\!\!=\!\!}\!\frac{e^{-4\pi kGb^2}}{e^{8\pi kGb\Delta}}
\!\propto\!e^{-\Delta/kT}
\eeq
Due to randomness of quantum decays, the average energy of a hawking mode emitted in one such spontaneous event is
\beq{}
\mathrm{for~fermions}:~\langle E\rangle\!=\!\frac{\hbar\omega e^{-\hbar\omega/kT}\!+\!0}{e^{-\hbar\omega/kT}\!+\!1}\!=\!\frac{\hbar\omega}{e^{\hbar\omega/kT}\!+\!1}
\eeq
\beq{}
\mathrm{for~bosons}:~\langle E\rangle\!=\!\frac{\sum_n n\hbar\omega e^{-n\hbar\omega/kT}}{\sum_n e^{-n\hbar\omega/kT}}\!=\!\frac{\hbar\omega}{e^{\hbar\omega/kT}\!-\!1}
\eeq
This is nothing but the spectrum of hawking radiations. However, we get it here basing on only standard quantum mechanics instead of any semi-classic consideration of quantum field theories in the curved background of space-time.

In fact, this simple quantum mechanic picture allows us to go more further than obtaining the power spectrum of hawking radiations. Basing on it, we may even construct a Hamiltonian thus explicitly unitary formulation for the whole process  as follows,
\beq{}
H=H_{\scriptscriptstyle\mathrm{BH}}+H_\mathrm{vac}+H_\mathrm{int}
\label{Hamiltonian0}
\eeq
\beq{}
=\left(\begin{matrix}b_{n}\\~\!\!&\!\!b_{n^\mm}\\~\!&\!~\!\!&\!\!\ddots
\\~&~&~&b_0\end{matrix}\right)
+\sum_k\hbar\omega_ka_k^\dagger a_k
+\sum_{|b_u\m\,b_v|}^{\hbar\omega_{\!k}=}\!\!g_{uv}b^\dagger_{uv}a_k
\label{Hamiltonian1}
\eeq
where $H_{\!_\mathrm{BH}}$, $H_\mathrm{vac}$ and $H_\mathrm{int}$ are respectively hamiltonians of the black hole, the environment and interactions between them two. The concrete form of $H_{\!_\mathrm{BH}}$ is unknown but unimportant. We need only to know that its eigenvalues are $\{b_n,b_{n^\mm},\cdots,b_1,b_0=0\}$ and respectively $\{w\!=\!e^{4k\pi Gb_n^2}$, $w^{\m\m}=\!e^{4k\pi Gb_{n^\mm}^2}$ $\cdots,1,1\}$-times degenerating. The vacuum hamiltonian $H_\mathrm{vac}$ is denoted by many harmonic oscillators, bosonic ones here for simplicity. When a mode's $\hbar\omega_k$ happens to be equal to the difference between some two eigenvalues of $H_{\!_\mathrm{BH}}$, it can go on shell and radiated away from the hole. The interaction part $H_\mathrm{int}$ between the holes and the environment functions to bring energies from the former to the latter and vice versa. Denoting the quantum state of a mass $b_\ell$, binding degree $u$ values in $\{1,2,\cdots,e^{4k\pi Gb_\ell^2}\}$, black hole and its environment consisting of hawking modes as $|\ell^u,n-\ell\rangle$, then\footnote{$b_u$, $b_v$ are abbreviations for $b_{\ell^{\!u}}=b_{\ell}$ and $b_{(\ell+k)^{\!v}}=b_{\ell+k}$ respectively.}
\beq{}
{\!}^{\,\hbar\omega_{\!k}\!=}_{b_{\!u}{\!\m\,}b_v\!}\!(b_{uv}^\dagger a_k)|\ell^v,n\m\ell\rangle=|(\ell+k)^u,n\m\ell\m k\rangle
\eeq
\beq{}
b_{uv}^\dagger\!=\!b_{vu}, a_{{\m}k}\!=\!a_{k}^\dagger
, g_{uv}\!=\!g_{uv}^*\!\propto\!\int\!\Psi^*_{\!\ell^v}\![r]\Psi_{\!(\ell+k)^{\!u}}\![r]dr
\label{guvDefinition}
\eeq
This transition $|\ell^v\!,n\m\ell\rangle\rightarrow|(\ell\!+\!k)^u\!,n\m\ell\m k\rangle$ could be induced by pure gravitational, especially mass monopole interactions. The proportionality in \eqref{guvDefinition} just reflects the fact that two black holes which have more similar microscopic wave functions could be more rapidly to transit to each other. If we further denote the initial state of the system as $|n^{\!w},0\rangle$, and the latter time state as follows
\beq{}
|\psi(t)\rangle\!=\!\sum_{\ell=0}^{n}\!\sum_{v=1}^{e^{S_\ell}}e^{{\m}ib_{\ell}t}\!c_{\ell^v\!}(t)|\ell^v,n\m\ell\rangle
\eeq
then the standard Schr\"odinger equation $i\hbar{\partial}_t\psi(t)=H\psi(t)$ will tell us
\bea{}
&&\hspace{-7mm}ih\bar\partial_tc_{\ell^v}\!(t)=(b_n-b_\ell)c_{\ell^v}+\sum_{j}^{\neq\ell}\sum_{u=1}^{e^{S_{\!j}}}g_{vu}e^{i(b_{\!\ell}-b_{\!j})t}c_{j^{\!u}}\!(t)
\\
&&\hspace{-7mm}c_{n^{\!w}}\!(0)\rule[2pt]{2mm}{0.5pt}1\!=\!c_{n_\m^u}(0)\!=\!c_{n_{\m\,\m}^v}(0)\!\cdots\!=\!0
\eea
If we know which binding state the initial black hole is at, i.e. the $w$ value in $n^{\!w}$, then we will be able to predict its latter time evolutions exactly, $n^{\!w}\xrightarrow{g_{wu}}n_{\m\m}^u\xrightarrow{g_{uv}}n_{\m\m\,\m\m}^v\cdots$. Conversely, if we can precisely monitor a black hole's evaporation process, especially the time feature of its horizon size variations, then we will be able to infer its initial state exactly. However, in semi classic discussion of hawking radiations, the initial state effects are completely averaged. All initial states are assumed to decay indistinguishably to the final state with equal chances, thus leading to the information missing puzzle.

In practical observations, initial states of black holes are almost unknowable. However, according to the general idea of probability theories, it is very natural that they have more chances to lie on positions of the phase space, binding degree space here, around which the density of microstate takes more larger values. Experiences from TABLE.\ref{shelling2mpl}, TABLE.\ref{shelling3mpl} and more larger black holes' microstate listing tells us that, such positions happens to be the classically  continuous distribution of matter contents inside the horizon. Comparing with the conventional black hole pictures with central singularity, the two body system in our pictures has totally different quadrupole  structures, so is dis/verifiable through gravitational wave observations, such as those reported in GW150914-170814 \cite{gw150914, gw151226, gw170104, gw170608, gw170814}. However, in all these observations, the inner structure of black holes are not measured because theoretical temples \cite{gwNumericA, gwNumericB, gwNumericC, gwNumericD, gwNumericE, gwNumericF, ligoTempleA, ligoTempleB, ligoTempleC, ligoTempleD} used in them to extract information from the highly noised signals adopt simple horizon boundary conditions to account for effects following from non-trivial quadrupole structures of the system. If we consider inner structure effects in theoretic temples, we expect to see that the gravitational wave forms, e.g. FIG.1 in \cite{gw150914}, following from binary black holes will become more close to that measured in the binary neutron star events, e.g. FIG.2 of GW170817 \cite{gw170817a}  .

{\em Summaries and discussion} Continuing our ideas in \cite{dfzeng2017}, we build a fuzzball picture for Scharzschild black holes and an explicitly unitary derivation for hawking radiations, thus a resolution to the information missing puzzles in this work. In our fuzzball pictures, the mass/energy contents of Scharzschild black holes are experiencing periodical oscillation across the central point with speed of light. Quantum mechanically, this oscillation happens only in some eigen-modes whose quantum wave function has maximal values only inside the horizon but is nonzero either outside it. By enumerating method, we show that the number of these eigen-modes happens to be exponentials of the Bekenstein-Hawking entropy.  In our derivations for the hawking radiation pow spectrum, the key is to average over the initial state's effects and to write the probability simply as the final state's degenerating multiplicity. While to answer the question on how the information is carried away by radiations, we construct a hamiltonian formulation for the whole process.

Many new works could be done from our fuzzball pictures. Most direct but non-trivially, generalises to other dimension, and to other asymptotic background such as AdS and dS space-times would be very interesting and necessary for better understanding of the origin of Bekenstein-Hawking entropy's area law feature. On the other hand, the most interesting work may be the experimental dis/verification of our inner structure picture for black holes through gravitation wave observation. Considering data accumulations today and the observation technique's maturity for this exploration \cite{gw150914, gw151226, gw170104, gw170608, gw170814, gw170817a}, what we need for this dis/verification work is just a theoretical temple to model the inner-structure carrying binary black hole's merging. We wish to come back this point in near futures.

\section*{Acknowledgements}
Thanks to colleagues are to be added when comments or suggestions are received. We thank NSFC and NSFB to give us chances say nothing here on financial supports.

\begin{table}
\bea{}
&&\begin{matrix}
\vspace{-6mm}\\
\hline
\{m_i\}/M_\mathrm{pl}&\{3\}&\{\frac{8}{3},\frac{1}{3}\}&\{\frac{7}{3},\frac{2}{3}\}
\\
M_i\!\equiv\!\sum m_{j\leqslant i}&\{3\}&\{\frac{8}{3},3\}&\{\frac{7}{3},3\}
\\
GM_im_i&\{9\}&\{\frac{64}{9},1\}&\{\frac{49}{9},1\}
\\
\frac{GM_im_i}{(1\leqslant)\beta}\in\mathbb{Z}\!&\!\{1,\!2\cddot9\}\!&\!\{(1,\!2\cddot7)(1)\}\!&\!\{(1,2\cdots5)(1)\}
\\
\#\mathrm{state}&9&7\cdot1&5\cdot1
\\
\hline
\end{matrix}\nonumber
\\
&&\begin{matrix}
\{m_i\}\!\!\!\!&\!\!\!\!\{\frac{6}{3},\frac{3}{3}\}&\{\frac{5}{3},\frac{4}{3}\}&\{\frac{4}{3},\frac{5}{3}\}&\{\frac{3}{3},\frac{6}{3}\}
\\
M_i\!\!\!\!&\!\!\!\!\{\frac{6}{3},3\}&\{\frac{5}{3},3\}&\{\frac{4}{3},3\}&\{\frac{3}{3},3\}
\\
GM_im_i\!\!\!\!&\!\!\!\!\{4,3\}&\{\frac{25}{9},4\}&\{\frac{16}{9},5\}&\{1,6\}
\\
&\hspace{-5mm}\{(1\cddot4)(1,\!2,\!3)\}\!\!&\!\!\{(1,\!2)(1\cddot4)\}\!\!&\!\!\{(1)(1\cddot5)\}\!\!&\!\!\{(1)(1\cddot6)\}
\\
\#\mathrm{state}\!\!\!\!&\!\!\!\!4\cdot3&2\cdot4&1\cdot5&1\cdot6
\\
\hline
\end{matrix}\nonumber
\\
&&\begin{matrix}
\{m_i\}\!\!&\!\!\{\frac{55}{24},\frac{3}{8},\frac{1}{3}\}\!&\!\{\frac{46}{24},\frac{6}{8},\frac{1}{3}\}\!&\!\{\frac{37}{24},\frac{9}{8},\frac{1}{3}\}\!&\!\{\frac{28}{24},\frac{12}{8},\frac{1}{3}\}
\\
M_i\!\!&\!\!\{\frac{55}{24},\frac{8}{3},3\}\!&\!\{\frac{46}{24},\frac{8}{3},3\}\!&\!\{\frac{37}{24},\frac{8}{3},3\}\!&\!\{\frac{28}{24},\frac{8}{3},3\}
\\
GM_im_i\!\!\!\!&\!\!\!\!\{5.,1,1\}\!&\!\{3.,2,1\}\!&\!\{2.,3,1\}\!&\!\{1.,4,1\}
\\
\cddot\!\!&\!\!\{(1\cddot5)\}\!&\!\{(123)(12)\}\!&\!\{(12)(123)\}\!&\!\{(1234)\}
\\
\#\mathrm{state}\!\!\!\!&\!\!\!\!5\!\cdot\!1\!\cdot\!1\!&\!3\!\cdot\!2\!\cdot\!1\!&\!2\!\cdot\!3\!\cdot\!1\!&\!1\!\cdot\!4\!\cdot\!1
\\
\hline
\end{matrix}\nonumber
\\
&&\begin{matrix}
\{m_i\}&\{\frac{40}{21},\frac{3}{7},\frac{2}{3}\}&\{\frac{31}{21},\frac{6}{7},\frac{2}{3}\}&\{\frac{22}{21},\frac{9}{7},\frac{2}{3}\}
\\
M_i&\{\frac{40}{21},\frac{7}{3},3\}&\{\frac{31}{21},\frac{7}{3},3\}&\{\frac{22}{21},\frac{7}{3},3\}
\\
GM_im_i&\{3.,1,2\}&\{2.,2,2\}&\{1.,3,2\}
\\
\frac{GM_im_i}{(1\leqslant)\beta}&\{(123)(12)\}&\{(12)(12)(12)\}&\{(123)(12)\}
\\
\#\mathrm{state}&3\!\cdot\!1\!\cdot\!2&2\!\cdot\!2\!\cdot\!2&1\!\cdot\!3\!\cdot\!2
\\
\hline
\end{matrix}\nonumber
\\
&&\begin{matrix}
\{m_i\}&\{\frac{3}{2},\frac{1}{2},\frac{3}{3}\}&\{1,\frac{2}{2},\frac{3}{3}\}&\{\frac{16}{15},\frac{3}{5},\frac{4}{3}\}
\\
M_i&\{\frac{3}{2},2,3\}&\{1,2,3\}&\{\frac{16}{15},\frac{5}{3},3\}
\\
GM_im_i&\{2.,1,3\}&\{1,2,3\}&\{1.,1,4\}
\\
\frac{GM_im_i}{(1\leqslant)\beta}&\{(12)(123)\}&\{(12)(123)\}&\{(1234)\}
\\
\#\mathrm{state}&2\!\cdot\!1\!\cdot\!3&1\!\cdot\!2\!\cdot\!3&1\!\cdot\!1\!\cdot\!4\rule{15mm}{0pt}
\\
\hline
\end{matrix}\nonumber
\\
&&\begin{matrix}
m_i\!\!\!&\!\!\!\{\frac{2449}{1320},\frac{24}{55},\frac{3}{8},\frac{1}{3}\}\!\!&\!\!\{\frac{1873}{1320},\frac{48}{55},\frac{3}{8},\frac{1}{3}\}\!\!&\!\!\{\frac{385}{276},\frac{24}{46},\frac{6}{8},\frac{1}{3}\}
\\
M_i\!\!\!&\!\!\!\{\frac{2449}{1320},\frac{55}{24},\frac{8}{3},3\}\!\!&\!\!\{\frac{1873}{1320},\frac{55}{24},\frac{8}{3},3\}\!\!&\!\!\{\frac{385}{276},\frac{46}{24},\frac{8}{3},3\}
\\
GM_im_i\!\!\!&\!\!\!\{3.,1,1,1\}\!&\!\{2.,2,1,1\}\!&\!\{1.,1,2,1\}
\\
\frac{GM_im_i}{(1\leqslant)\beta}\!\!\!&\!\!\!\{(123)\}\!&\!\{(12)(12)\}\!&\!\{(12)\}
\\
\#\mathrm{state}\!\!\!&\!\!\!3\!\cdot\!1\!\cdot\!1\!\cdot\!1\!&\!2\!\cdot\!2\!\cdot\!1\!\cdot\!1\!&\!1\!\cdot\!1\!\cdot\!2\!\cdot\!1
\\
\hline
\end{matrix}\nonumber
\\
&&\begin{matrix}
\{m_i\}&\{\frac{1159}{840},\frac{21}{40},\frac{3}{7},\frac{2}{3}\}&\{\frac{4255201}{3232680},\frac{1320}{2449},\frac{24}{55},\frac{3}{8},\frac{1}{3}\}
\\
M_i&\{\frac{1159}{840},\frac{40}{21},\frac{7}{3},3\}&\{\frac{4255201}{3232680},\frac{2449}{1320},\frac{55}{24},\frac{8}{3},3\}
\\
GM_im_i&\{1.,1,1,2\}&\{1.,1,1,1,1\}
\\
\frac{GM_im_i}{(1\leqslant)\beta}&\{(1)\cdots(1,\!2)\}&\{(1)(1)\cdots\}
\\
\#\mathrm{state}&1\!\cdot\!1\!\cdot\!1\!\cdot\!1\!\cdot\!2&1\!\cdot\!1\!\cdot\!1\!\cdot\!1\!\cdot\!1
\\
\hline
\end{matrix}\nonumber
\eea
\caption{The same as TABLE \ref{shelling2mpl}, but for a $3M_\mathrm{pl}$ mass black hole. Other layering schemes such as decomposing $3M_\mathrm{pl}$ into $\{\frac{8M_\mathrm{pl}}{3}-\epsilon$, $\frac{M_\mathrm{pl}}{3}+\epsilon\}$ with $|\epsilon|<\frac{M_\mathrm{pl}}{3}$ will not lead to quantum numbers different from listed above. So this table lists out all possible quantum state of a $3M_\mathrm{pl}$ mass Schwarzschild black hole exclusively, whose total number is $126$.}
\label{shelling3mpl}
\end{table}


\begin{thebibliography}{99}

\bibitem{cardoso2017}
V. Cardoso, P. Pani,
``Tests for the existence of horizons through gravitational wave echoes'',
{\em Nature Astronomy} {\bf1:} 586-591 (2017),
\href{https://arxiv.org/abs/1709.01525}{arXiv: 1709.01525}.

\bibitem{mathur2009}
S. D. Mathur,
``The information paradox: A pedagogical introduction'',
{\em 	Class.Quant.Grav.} {\bf 26:} 224001(2009),
\href{https://arxiv.org/abs/0909.1038}{arXiv: 0909.1038}

\bibitem{hawking1976}
S. Hawking,
``Breakdown of Predictability in Gravitational Collapse'',
{\em Phys. Rev. } {\bf D14, } 2460 (1976)

\bibitem{hawking1975cmp}
S. Hawking,
``Black hole explosions'', {\em Nature} {\bf248} (1974) 30;
``Particle creation by black holes'',
{\em Comm. Math. Phys.} {\bf43} (1975) 199.

\bibitem{Mathur0909}
S. D. Mathur,
``The Information paradox: A Pedagogical introduction'',
{\em Class. Quant. Grav.} {\bf26} (2009) 224001,
\href{http://arxiv.org/abs/arXiv:0909.1038}{arXiv:0909.1038}

\bibitem{polchinski2016}
J. Polchinski,
``The Black Hole Information Problem'',
{\em New Frontiers in Fields and Strings}, TASI 2015, chapter 6,
\href{https://arxiv.org/abs/1609.04036}{arXiv: 1609.04036}.

 \bibitem{maldacena1997}
 J.~M. Maldacena,
 ``The large N limit of superconformal field theories and supergravity'',
 {\em Adv. Theor. Math. Phys.} {\bf 2} (1998) 231--252,
 \href{http://xxx.lanl.gov/abs/hep-th/9711200}{arXiv: hep-th/9711200};
 
\bibitem{DasMathur9601}
S. R. Das, S. D. Mathur,
``Excitations of D-strings, Entropy and Duality''
{\em Phys. Lett.} {\bf B375} (1996) 103,
\href{https://arxiv.org/abs/hep-th/9601152}{arXiv: hep-th/9601152}

\bibitem{MaldacenaSusskind9604}
J. M. Maldacena, L. Susskind,
``D-branes and Fat Black Holes'',
{\em Nucl. Phys.} {\bf B475} (1996) 679,
\href{https://arxiv.org/abs/hep-th/9604042}{arXiv:  hep-th/9604042}.

\bibitem{StromingVafa9601}
A. Strominger, C. Vafa,
``Microscopic Origin of the Bekenstein-Hawking Entropy''
{\em Phys. Lett} {\bf B379} (1996) 99,
\href{https://arxiv.org/abs/hep-th/9601029}{arXiv: 9601029}.

\bibitem{Mathur9706}
S. D. Mathur,
``Emission rates, the Correspondence Principle and the Information Paradox'',
{\em Nucl. Phys.}{\bf B529} (1998) 295,
\href{https://arxiv.org/abs/hep-th/9706151}{arXiv: hep-th/9706151}.
 
\bibitem{LuninMathur0109}
O. Lunin, S. D. Mathur,
``AdS/CFT duality and the black hole information paradox'',
{\em Nucl. Phys.}{\bf B623} (2002) 342,
\href{https://arxiv.org/abs/hep-th/0109154}{arXiv: hep-th/0109154}.

\bibitem{LuninMathur0202}
O. Lunin, S. D. Mathur,
``Statistical interpretation of Bekenstein entropy for systems with a stretched horizon'',
{\em Phys. Rev. Lett.} {\bf 88} (2002) 211303,
\href{https://arxiv.org/abs/hep-th/0202072}{arXiv: hep-th/0202072}.

\bibitem{LuninMaldacenaMaoz0212}
O. Lunin, J. Maldacena, L. Maoz,
``Gravity solutions for the D1-D5 system with angular momentum'',
\href{https://arxiv.org/abs/hep-th/0212210}{arXiv: hep-th/0212210}.

\bibitem{MT0103}
D. Mateos, P. K. Townsend,	
``Supertubes'',
{\em Phys. Rev. Lett.} {\bf87} (2001) 011602,
\href{http://arxiv.org/abs/hep-th/0103030}{arXiv: hep-th/0103030}

\bibitem{GMathur0412}
S. Giusto, S. D. Mathur,
``Fuzzball geometries and higher derivative corrections for extremal holes'',
{\em Nucl. Phys.} {\bf B738} (2006) 48-75,
\href{http://arxiv.org/abs/hep-th/0412133}{arXiv: hep-th/0412133}

\bibitem{mathur0502}
S. D. Mathur,
``The Fuzzball proposal for black holes: An Elementary review'',
{\it Fortsch.Phys.} {\bf53} (2005) 793-827,
\href{http://arxiv.org/abs/hep-th/0502050}{arXiv: 0502050}.

\bibitem{Jejjala0504}
``Non-supersymmetric smooth geometries and D1-D5-P bound states'',
V. Jejjala, O. Madden, S. F. Ross, G. Titchener,
{\em Phys. Rev.} {\bf D71} (2005) 124030,
\href{http://arxiv.org/abs/hep-th/0504181}{arXiv: hep-th/0504181}.

\bibitem{KST0704}
I. Kanitscheider, K. Skenderis, M. Taylor,
``Fuzzballs with internal excitations'',
{\em JHEP} {\bf 0706} (2007) 056,
\href{http://arxiv.org/abs/arXiv:0704.0690}{arXiv: 0704.0690}

\bibitem{Mathur0706}
S. D. Mathur,
``Black hole size and phase space volumes',
\href{http://arxiv.org/abs/arXiv:0706.3884}{arXiv:0706.3884}
 
\bibitem{blackholewar}
L. Susskind,
``The Black Hole War: My Battle with Stephen Hawking to Make the World Safe for Quantum Mechanics'',
Hachette Inc. ISBN 978-0-316-01640-7.
 
\bibitem{fireworksAMPS2012}
A. Almheiri, D. Marolf and J. Polchinski et al,
``Black Holes: Complementarity or Firewalls?''
{\it JHEP} {\bf1302} (2013) 062,
\href{http://arxiv.org/abs/1207.3123}{arXiv:1207.3123}

\bibitem{fireworksAMPS2013}
A. Almheiri, D. Marolf and J. Polchinski et al, 
``An Apologia for Firewalls'', 
{\it JHEP} {\bf1309}(2013) 018,
\href{http://arxiv.org/abs/1304.6483}{arXiv:1304.6483}.
 
\bibitem{EREPR}
J. Maldacena, L. Susskind,
``Cool horizons for entangled black holes'',
{\em Fortschr. Phys.} {\bf20130020} (2013) 1,
\href{https://arxiv.org/abs/1306.0533}{arXiv: 1306.0533}.
 
\bibitem{stojkovic1401}
A. Saini, D. Stojkovic,
``Nonlocal (but also nonsingular) physics at the last stages of gravitational collapse'',
{\it Phys. Rev.} {\bf D89} (2014), 044003,
\href{https://128.84.21.199/abs/1401.6182}{arXiv: 1401.6182}.

\bibitem{stojkovic1503}
A. Saini, D. Stojkovic,
``Radiation from a collapsing object is manifestly unitary'',
{\it Phys. Rev. Lett.} {\bf114} (2015), 111301,
\href{https://128.84.21.199/abs/1503.01487v3}{arXiv: 1503.01487}.

\bibitem{stojkovic1601}
D.-C. Dai and D. Stojkovic, 
``Pre-Hawking radiation may allow for reconstruction of the mass distribution of the collapsing object'', {\em Phys. Lett.} {\bf B758} (2016) 429
\href{https://arxiv.org/abs/1601.07921}{arXiv: 1601.07921}

\bibitem{bardeen1706}
J. Bardeen,
``The semi-classical stress-energy tensor in a Schwarzschild background, the information paradox, and the fate of an evaporating black hole'',
\href{https://arxiv.org/abs/1706.09204}{arXiv: 1706.09204}.

\bibitem{halyo1706}
E. Halyo,
``The Holographic Entanglement Entropy of Schwarzschild Black Holes'',
\href{https://arxiv.org/abs/1706.07428}{arXiv: 1706.07428}

\bibitem{stojkovic0609}
T. Vachaspati, D. Stojkovic and L.M. Krauss, 
``Observation of incipient black holes and the information loss problem'', 
{\em Phys. Rev.} {\bf D76} (2007) 024005,
\href{https://arxiv.org/abs/gr-qc/0609024}{arXiv: gr-qc/0609024}

\bibitem{sunYi1102}
Yi Sun,
``Black hole: Never forms, or never evaporates'',
{\em JCAP} {\bf 1101} (2011) 031,
\href{http://arxiv.org/abs/arXiv:1102.2609}{arXiv: 1102.2609}

\bibitem{KMY1302}
H. Kawai, Y. Matsuo, Y. Yokokura,
``A Self-consistent Model of the Black Hole Evaporation ''
{\em Int. J. Mod. Phys.} {\bf A28} (2013) 1350050,
\href{http://arxiv.org/abs/arXiv:1302.4733}{arXiv: 1302.4733}

\bibitem{zhangBc1305}
B.-c. Zhang et al, 
``Information conservation is fundamental: Recovering the lost information in Hawking radiation''，{\it Int. Journal of Mod. Phys.} {\bf D22} (2013)1341014,
\href{http://arxiv.org/abs/arXiv:1305.6341}{arXiv: 1305.6341}.

\bibitem{caiQingyu1608}
Q.-y. Cai, C.-p. Sun, L. You, 
``Information-carrying Hawking radiation and the number of microstate for a black hole''，
{\it Nucl. Phys.} {\bf B905} (2016) 327,
\href{http://arxiv.org/abs/arXiv:1608.08304}{arXiv: 1608.08304}.

\bibitem{hopeiming1609}
Pei-ming Ho,
``Asymptotic black holes'',
{\em Class. and Quan. Grav.} {\bf vol34} (2017) 085006,
\href{http://arxiv.org/abs/arXiv:1609.05775}{arXiv: 1609.05775}.

\bibitem{hawking2016}
``Soft Hair on Black Holes''
S. Hawking, M. Perry, A. Strominger,
{\it Phys. Rev. Lett.} {\bf116} (2016) 231301,
\href{http://arxiv.org/abs/1601.00921}{arXiv:1601.00921}
 
\bibitem{bousso1706}
R. Bousso, M. Porrati,
``Soft Hair as a Soft Wig'',
\href{https://arxiv.org/abs/1706.00436}{arXiv: 1706.00436}.

\bibitem{giddings1706}
W. Donnelly, S. Giddings,
``How is quantum information localized in gravity?'',
{\em Phys. Rev.} {D96} (2017) 086013,
\href{https://arxiv.org/abs/1706.03104}{arXiv: 1706.03104} 

\bibitem{dfzeng2017}
Ding-fang Zeng, 
``Resolving the Schwarzschild singularity in both classic and quantum gravities'',
{\em Nucl. Phys.} {\b917} 178-192.  

\bibitem{Yodzis1973}
P. Yodzis, H.J. Seifert and H. Müller zum Hagen,
``On the occurrence of naked singularities in general relativity'',
{\it Commun. Math. Phys.} {\bf34} (1973), 135.

\bibitem{Yodzis1974}
P. Yodzis, H.J. Seifert and H. Müller zum Hagen,
``On the occurrence of naked singularities in general relativity. II'',
{\it Commun. Math. Phys.} {\bf37} (1974), 29.

\bibitem{GTV2012}
D. M. Gitman, I. V. Tyutin, B. L. Voronov,
``Self-adjoint Extentions in Quantum Mechanics, General Theory and Applications to Shrodinger and Dirac Equations with Singular Potentials'', Sec. 7.2,
{\em Progress in Mathematical Physics}, {\bf vol 62}, Birkhäuser Boston (2012).
\href{http://www.springer.com/us/book/9780817644000}{DOI: 10.1007/978-0-8176-4662-2}

\bibitem{gw150914}
The LIGO Scientific Collaboration, the Virgo Collaboration,
``Observation of Gravitational Waves from a Binary Black Hole Merger'',
{\it Phys. Rev. Lett.} {\bf116} (2016) 061102,
\href{https://arxiv.org/abs/1602.03837}{arXiv:1602.03837}.

\bibitem{gw151226}	
LIGO Scientific and Virgo Collaborations,
``GW151226: Observation of Gravitational Waves from a 22-Solar-Mass Binary Black Hole Coalescence'',
{\it Phys. Rev. Lett.} {\bf116} (2016) 241103,
\href{http://arxiv.org/abs/arXiv:1606.04855}{arXiv:1606.04855}

\bibitem{gw170104}
LIGO Scientific and Virgo Collaborations, 
"GW170104: Observation of a 50-Solar-Mass Binary Black Hole Coalescence at Redshift 0.2", 
{\it Phys. Rev. Lett.} {\bf118} (2017) 221101,
\href{http://arxiv.org/abs/arXiv:1706.01812}{arXiv:1706.01812}.

\bibitem{gw170608}
LIGO Scientific and Virgo Collaborations, 
"GW170608: Observation of a 19-solar-mass Binary Black Hole Coalescence",
{\it The Astro. Journal. Lett.} 851 (2),
\href{http://arxiv.org/abs/arXiv:1711.05578}{arXiv:1711.05578}

\bibitem{gw170814} 
LIGO Scientific and Virgo Collaborations, 
"GW170814: A three-detector observation of gravitational waves from a binary black hole coalescence", {\it Phys. Rev. Lett.} {\bf119} (2017) 141101
\href{http://arxiv.org/abs/arXiv:1709.09660}{arXiv:1709.09660}.


\bibitem{gwNumericA}
L. Blanchet et al,
``Gravitational-Radiation Damping of Compact Binary Systems to Second Post-Newtonian Order'',
{\it Phys. Rev. Lett.} {\bf74} (1995) 3515,
\href{http://arxiv.org/abs/gr-qc/9501027}{arXiv:gr-qc/9501017}

\bibitem{gwNumericB}
``Gravitational Radiation from Post-Newtonian Sources and Inspiralling Compact Binaries'',
{\it Liv. Rev. Rel.} {\bf17} (2014) 2.
\href{http://arxiv.org/abs/arXiv:1310.1528}{arXiv:1310.1528}.

\bibitem{gwNumericC}
A. Buonanno and T. Damour,
``Effective one-body approach to general relativistic two-body dynamics'',
{\it Phys. Rev.} {\bf D59} {1999} 084006,
\href{http://arxiv.org/abs/gr-qc/9811091}{arXiv:gr-qc/9811091}.

\bibitem{gwNumericD}
F. Pretorius,
``Evolution of Binary Black-Hole Spacetimes'',
{\it Phys. Rev. Lett.} {\bf95} (2005) 121101,
\href{http://arxiv.org/abs/gr-qc/0507014}{arXiv: gr-qc/0507014}.

\bibitem{gwNumericE}
M. Campanelli et al,
``Accurate Evolutions of Orbiting Black-Hole Binaries without Excision'',
{\it Phys. Rev. Lett.} {\bf96} (2006) 111101,
\href{http://arxiv.org/abs/gr-qc/0511048}{arXiv: gr-qc/0511048}

\bibitem{gwNumericF}
J. G. Baker et al,
``Gravitational-Wave Extraction from an Inspiraling Configuration of Merging Black Holes'',
{\it Phys. Rev. Lett.} {\bf96} (2006) 111102,
\href{http://arxiv.org/abs/gr-qc/0511103}{arXiv: gr-qc/0511103}.

\bibitem{ligoTempleA}
A. Buonanno \& T. Damour,
``Transition from inspiral to plunge in binary black hole coalescences'',
{\it Phys. Rev.} {\bf D62} (2000) 064015,
\href{http://arxiv.org/abs/gr-qc/0001013}{arXiv:0001013}

\bibitem{ligoTempleB}
L. Blanchet, T. Damour, G. Esposito-Farèse \& B. R. Iyer,
``Gravitational Radiation from Inspiralling Compact Binaries Completed at the Third Post-Newtonian Order'',
{\it Phys. Rev. Lett.} {\bf93} (2004) 091101,
\href{http://arxiv.org/abs/gr-qc/0406012}{arXiv: gr-qc/0406012}

\bibitem{ligoTempleC}
A. Taracchini et al,
``Effective-one-body model for black-hole binaries with generic mass ratios and spins'',
{\it Phys. Rev.} {\bf D89} (2014) 061502,
\href{http://arxiv.org/abs/arXiv:1311.2544}{arXiv:1311.2544}.

\bibitem{ligoTempleD}
M. P\"urrer,
``Frequency-domain reduced order models for gravitational waves from aligned-spin compact binaries'',
{\it Class. Quan. Grav.} {\bf31} (2014) no.19, 195010.
\href{http://arxiv.org/abs/arXiv:1402.4146}{arXiv:1402.4146}.
 
\bibitem{gw170817a}
Abbott, B. P.  et al. (LIGO Scientific Collaboration \& Virgo Collaboration),
"GW170817: Observation of Gravitational Waves from a Binary Neutron Star Inspiral",
{\it Phys. Rev. Lett.} {\bf119} (2017) 161101,
\href{https://arxiv.org/abs/1710.05832}{arXiv:1710.05832 }

 \end{thebibliography}
\end{document}